\documentclass[journal]{IEEEtran}

\usepackage{cite}

\ifCLASSINFOpdf
  \usepackage[pdftex]{graphicx}

  \DeclareGraphicsExtensions{.pdf,.jpeg,.png}
\else

\fi

\usepackage{amsmath}
\usepackage{stix}
\usepackage{threeparttable}
\usepackage{array}
\usepackage{xcolor}
\usepackage{soul}
\soulregister\cite7
\usepackage{pdfpages}

\begin{document}
\twocolumn[
\begin{@twocolumnfalse}

\textbf{Author's pre-print} \\

\copyright 2021 IEEE. Personal use of this material is permitted. Permission from IEEE must be obtained for all other users, including reprinting/ republishing this material for advertising or promotional purposes, creating new collective works for resale or redistribution to servers or lists, or reuse of any copyrighted components of this work in other works.\\


\end{@twocolumnfalse}
]

\newpage
%
\title{Antenna Optimization for WBAN Based on Spherical Wave Functions De-Embedding}
%
%
%
\author{Lukas~Berkelmann, \IEEEmembership{Graduate Student Member, IEEE}, Hendrik~J\"aschke, Leonardo~M\"orlein, \IEEEmembership{Graduate Student Member, IEEE}, Lukas~Grundmann, \IEEEmembership{Graduate Student Member, IEEE}, and  Dirk~Manteuffel, \IEEEmembership{Member, IEEE}
\thanks{This work was supported by the Deutsche
Forschungsgemeinschaft (DFG) under grant MA 4981/11-1.
}
\thanks{The authors are with the Institute of Microwave and Wireless Systems, Leibniz University Hannover, Appelstr. 9A, 30167 Hannover, Germany \mbox{(e-mail:} \mbox{berkelmann@imw.uni-hannover.de;} \mbox{h.jaeschke@stud.uni-hannover.de;} \mbox{moerlein@imw.uni-hannover.de;} \mbox{grundmann@imw.uni-hannover.de;} \mbox{manteuffel@imw.uni-hannover.de)}
 (Hendrik J\"aschke, Leonardo M\"orlein and Lukas Grundmann contributed equally to this work.) (Corresponding author: Lukas Berkelmann)
}}

\maketitle

\begin{abstract}
Antennas for wireless body area networks (WBAN) need to be modeled with adapted methods because the coupling with the body tissue does not allow for a clear separation between antenna and channel. Especially for dynamically varying on-body channels due to changing body poses, e.g. with head-worn antennas, modeling is challenging and design goals for optimal antennas are difficult to determine. Therefore, in this paper, the modeling of WBAN channels using spherical wave functions (SWF) is utilized for antenna de-embedding and for deriving optimal antenna characteristics that maximize the transmission coefficient for the respective channel. It is evaluated how typical factors influencing WBAN channels (different body anatomies, body postures, and varying positions of the communication nodes), can be modeled statistically with SWF. An optimized antenna design is developed based on the derived optimization method, specifically adapted to the channel of on-body links with eye-wear applications. The results with the optimized antenna are compared to other standard antenna designs and validated against measurements.
\end{abstract}


%
\IEEEpeerreviewmaketitle

\section{Introduction}

\begin{figure}
    \centering
    \includegraphics[width=0.99\columnwidth]{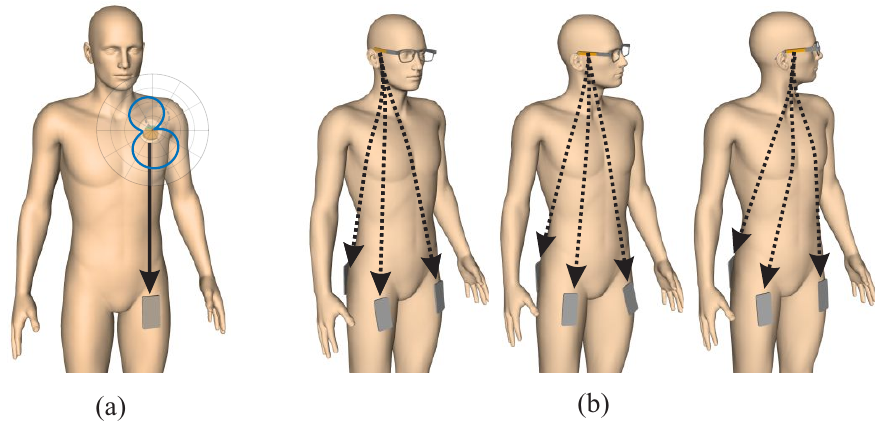}
    \caption{On-body channel modeling: (a) Static (e.g. implanted pacemaker); (b) Frame-wise dynamic (e.g. head-worn device)}
    \label{fig:DD_WBAN_application}
\end{figure}

\IEEEPARstart{W}{ireless} body area networks (WBAN) consist of computing devices in the vicinity of the body and are established in many areas. From the perspective of standard antenna modeling, the antennas in WBAN are embedded in the channel due to the coupling between the antennas and the human body.
Therefore, without antenna de-embedding, WBAN systems can only be characterized as a whole including transmitter, on-body channel, and receiver. By observing different antenna positions, and body poses, statistical path gain models and channel models can be implemented \cite{Gallo2011,Kumpuniemi2017,Naganawa2015,Uusitupa2013,Wang2009,Ali2015a,AkhoondzadehAsl2013}. However, the characterized channels are always specific to the given antenna type, antenna position and body pose. Thus, these models lack generality, which makes antenna optimization difficult.
For some WBAN applications, the on-body channel can be considered static, meaning that its overall propagation behavior is not significantly affected by body poses, etc. As an example, in Fig. \ref{fig:DD_WBAN_application}(a), the RF link between an implanted pacemaker and a smartphone located at the pocket of the user is depicted. It can be assumed that the main propagation channel on the torso is not significantly affected by different postures of e.g. the head or the legs. In this case, double-directional (DD) channel modeling can be used \cite{Steinbauer2001}, where the channel is modeled in dependency of the angular radiation properties at both transmitter and receiver. However, with WBAN, the issue of near-field interaction between antenna and body tissue must be resolved to separate the antennas from the channel. Therefore, we have recently proposed an approach of adapted on-body antenna parameters, such as an angular on-body gain pattern for quantifying an antenna's ability of exciting surface waves or creeping waves \cite{Berkelmann2021,Grimm2015,Grimm2014}.
On-body DD channel models can be found mainly for simplified geometries.
For example, Kamersgaard~et~al. have defined a comprehensive creeping wave channel model for ear-to-ear propagation based on elliptical trajectories around the head \cite{Kammersgaard2019}. The on-body gain can also be measured for real-world applications by using specifically designed antenna test ranges, as we have shown recently \cite{Berkelmann2021a}. This makes it an ideal measure for antenna optimization, where the optimization goal is to maximize the on-body gain in direction of the main propagation path.

However, with other WBAN applications, the propagation behavior of the on-body channel changes dynamically due to body posing. Thus, dominant paths of the propagation channel vary with body posture and a particular direction for optimizing the on-body gain cannot be determined. As an example, in Fig. \ref{fig:DD_WBAN_application}(b), an application with a head-worn device is depicted. Here, the polarization of the head-mounted antenna with respect to the torso and the other antennas is affected significantly if the head is turned. Furthermore, the location of the devices, e.g. of the smartphone as depicted in Fig. \ref{fig:DD_WBAN_application}(b), is often not known exactly. Current solutions for these cases based on empirical statistical channel models are, as already discussed above, always specific to the given antennas, since these are embedded in the channel. Therefore, they do not allow insight into the design of optimal antennas.

An inspiring approach for deriving antenna-unspecific channel models for WBAN has been proposed by Naganawa et al. \cite{Naganawa2017} using spherical wave functions (SWF).
Here, the channel is characterized by the coupling between all SWF at the source and the receiver. Similarly, the antennas can be characterized by their ability to excite those SWF.
Since the method can be implemented numerically, no simplification of the geometry is necessary. {However, the proposed SWF antenna characterization requires a multi-step calculation process to account for the near-field coupling of the antenna with the body tissue.}
As shown in \cite{Naganawa2017}, dynamic WBAN channels due to body posing as depicted in Fig.~\ref{fig:DD_WBAN_application} can be approached by frame-wise modeling of several static snapshots\cite{Naganawa2017}. 
Apart from WBAN applications, Arai et. al have recently utilized SWF modeling for antenna optimization to maximize MIMO channel capacities \cite{Arai2017}.

In this contribution, we investigate how SWF modeling can be used to determine optimal characteristics for WBAN antennas directly from the respective SWF channel models. Based thereon, optimized antenna designs adapted to the WBAN channel can be derived.
In Section \ref{sec:SWF_general}, the general SWF modeling approach is summarized. {The issue of the SWF antenna characterization, respectively the antenna de-embedding, while accounting for the near-field coupling to the body tissue is approached in Section {\ref{sec:SWF_WBAN}}. Compared to~\mbox{\cite{Naganawa2017}}, we propose a more rigorous and straightforward approach that can be implemented with standard EM modeling software.} In Section~\ref{sec:SWF_opt} an optimization method is derived, that enables the calculation of SWF-coefficients for maximizing the transmission coefficient in dependency of the SWF channel model. Furthermore, it is shown how the on-body radiation pattern of the optimal antenna for a certain channel can be obtained, which can serve as a guideline for optimized antenna designs.
In Section~\ref{sec:application}, the developed methods are applied and evaluated for the design of antennas for eye-wear applications as an example. Finally, in Section~\ref{sec:measurements} the results are validated against measurements. Additionally, a possible application of SWF modeling for statistical evaluation and optimization of WBAN antenna designs based on a key performance indicator (KPI) is evaluated.

\section{SWF Channel Modeling and Antenna De-Embedding}
\label{sec:SWF_general}
\begin{figure}
    \centering
    \includegraphics[width=0.99\columnwidth]{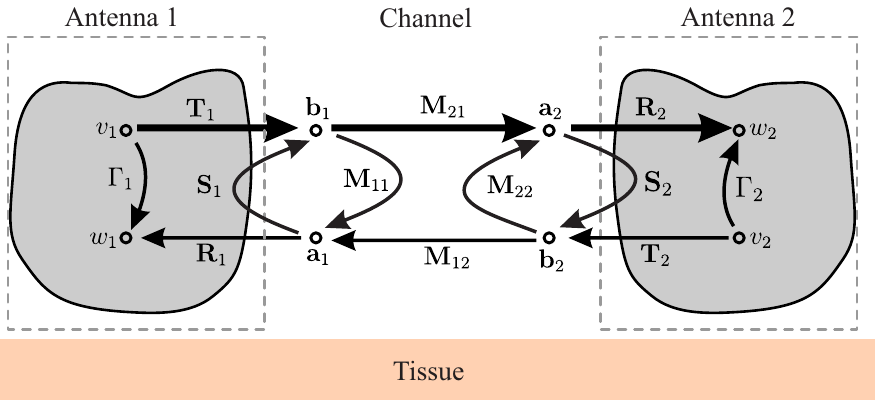}
    \caption{Antenna coupling between two antennas in presence of body tissue. The SWF scattering matrices~$\mathbf{M}_{mn}$ describe the mode-to-mode transmission and reflection behavior of the channel. Antennas are characterized by their SWF transmission and receiving behavior denoted by the vectors~$\mathbf{T}$ and~$\mathbf{R}$. }
    \label{fig:SWF_Flow_graph_complete}
\end{figure}

Spherical wave functions (SWF) represent a complete and orthogonal set of basis functions for solutions to Maxwell's equations.
Any electrical field can be decomposed into a linear combination of SWF at an arbitrary origin point by \cite{Hansen1988}:
\begin{equation} \label{eq:e_total_sum}
    \mathbf{E} = k \sqrt{\eta} \sum_{j=1}^J \left( b_j\mathbf{F}_j^{(4)} + a_j \mathbf{F}_j^{(3)} \right),
\end{equation}
whereby~$k$ is the wavenumber,~$\eta$ is the wave impedance in free-space, and~$j$ is the mode index.~$\mathbf{F}_j^{(4)}$ represent outgoing SWF and~$b_j\in\mathbb{C}$ are the associated weighting coefficients. Accordingly,~$\mathbf{F}_j^{(3)}$ and~$a_j\in\mathbb{C}$ represent incoming SWF.\footnote{In contrast to \cite{Hansen1988}, we assume the time-dependence~$e^{j\omega t}$.} Theoretically, an infinite number of SWF modes $J$ is necessary for the description of arbitrary fields. However, it is well known that fields radiated by antennas can practically be characterized by a finite number of modes due to the radial cutoff property of~$\mathbf{F}_j$\cite{Hansen1988}. An estimation for the truncation of~$j$ is usually determined by the radius of the minimum sphere enclosing the whole antenna structure \cite{Hansen1988}.

As shown by Pirkl, the coupling between two antennas in arbitrary environments can be described utilizing spherical wave scattering matrices \cite{Pirkl2012}. If this is applied to WBAN, the channel is partly formed by body tissue, as depicted in Fig.~\ref{fig:SWF_Flow_graph_complete}.
If the link between a transmitting antenna 1 and a receiving antenna 2 is to be calculated, the transmission \mbox{vector~$\mathbf{T}_1\in\mathbb{C}^{J\times 1}$} quantifies the transitional behavior between the incident wave~$v_1$ at the physical port~1, and the vector~\mbox{$\mathbf{b}_1\in\mathbb{C}^{ J\times1}$} consisting of all outgoing SWF coefficient~$b_j$ at antenna~1 (assuming~$\mathbf{a}_1=\mathbf{0}$) as:
\begin{equation}
\label{eq:T}
   \mathbf{b}_1=\mathbf{T}_1v_1\,. 
\end{equation}
$\mathbf{T}_1$ can be calculated numerically from the radiated fields of the transmitting antenna by assuming only coefficients for outgoing~waves~$b_j$ in~\eqref{eq:e_total_sum}.
The receiving \mbox{vector~$\mathbf{R}_2\in\mathbb{C}^{1\times J}$} of antenna~2 can be defined equivalently  (assuming~$\mathbf{b}_2=\mathbf{0}$) as:
\begin{equation}
   w_2=\mathbf{R}_2\mathbf{a}_2,
\end{equation}
with~$w_2$ the outgoing wave at the physical port~2 and \mbox{$\mathbf{a}_2\in\mathbb{C}^{ J\times1}$} the SWF coefficient vector consisting of the coefficients~$a_j$ of all incoming SWF. For reciprocal antennas,~$\mathbf{R}_2$ can be calculated from~$\mathbf{T}_2$ and vice versa \cite{Hansen1988}.
Incoming waves at the antennas are also partially re-radiated/backscattered, which is quantified e.g for antenna~1 by the scattering matrix~\mbox{$\mathbf{S}_1\in\mathbb{C}^{J\times J}$}. Finally, the free-space input reflection coefficient at the physical port of e.g. antenna~1. is defined as~$\Gamma_1$. 

SWF channel modeling is based on the SWF scattering matrices~$\mathbf{M}_{mn}\in\mathbb{C}^{J\times J}$, ref.~Fig.~\ref{fig:SWF_Flow_graph_complete}. For example, the transmission from antenna~1 in form of the outgoing spherical waves~$\mathbf{b}_1$ to the incoming waves~$\mathbf{a}_2$ at antenna~2 is described by (assuming~$\mathbf{b}_2=0$):
\begin{equation}
\label{eq:M}
    \mathbf{a}_\mathrm{2} = \mathbf{M}_{21} \, \mathbf{b}_\mathrm{1}.
\end{equation}
{Based on SWF modeling, the near-field coupling of the antennas with the body tissue can be described as backscattering of the channel and can be characterized by:}
\begin{equation}
\label{eq:a}
    \mathbf{a}_1=\mathbf{M}_{11}\mathbf{b}_1.
\end{equation}
As shown by Pirkl, the SWF scattering matrices of the channel can be calculated numerically \cite{Pirkl2012}. 
In the most simple case, with ideally matched~antennas \mbox{($\Gamma=0$)} and a channel with negligible back~scattering \mbox{($\mathbf{M}_{11}$=\,$\mathbf{M}_{22}\approx \mathbf{0}$)}, the narrow-band transmission coefficient~$S_{21}$ can be calculated as \cite{Glazunov2010}:
\begin{equation}
\label{eq:S21_FS}
    S_{21}=w_2v_1^{-1}=\mathbf{R_2\,M_{21}\,T_1}. 
\end{equation}
As can be seen from \eqref{eq:S21_FS}, the system is now separated into three building blocks. Once the channel is characterized with~$\mathbf{M}_{21}$, the transmission coefficient with any combination of antennas~$\mathbf{R}_{2}$,~$\mathbf{T}_{1}$ can be calculated by simply carrying out the matrix multiplication in \eqref{eq:S21_FS}. Thus, for optimizing the antenna in the scenario depicted in Fig.~\ref{fig:DD_WBAN_application}(b), the channel matrices to all possible receiver locations in all body poses to be considered can be calculated first. Then, the link to all considered receiving locations in different body poses can be calculated using \eqref{eq:S21_FS} with every design iteration from a single simulation of the transmitting antenna by recalculating $\mathbf{T}_1$.

\section{SWF Antenna De-Embedding in Presence of a Backscatterer}
\label{sec:SWF_WBAN}
{A difficulty that arises specifically when applying SWF antenna de-embedding for WBAN is that due to the near-field coupling of the antennas with the tissue, the assumption \mbox{($\mathbf{M}_{11}$=\,$\mathbf{M}_{22}\approx \mathbf{0}$)} is not valid.} 
Therefore, the system~response~\eqref{eq:S21_FS} becomes \cite{Pirkl2012}:
\begin{equation}
\label{eq:S21_OB_org}
    S_{21}=\mathbf{R}_2\underbrace{\left(\mathbf{I}-\mathbf{M}_{22}\mathbf{S}_2\right)^{-1}}_{\mathbf{K}_2}\mathbf{M}_{21} \underbrace{\left(\mathbf{I}-\mathbf{S}_1\mathbf{M}_{11}\right)^{-1}}_{\mathbf{K}_1}\mathbf{T}_1,
\end{equation}
where the matrices~${\mathbf{K}_1}$ and~${\mathbf{K}_2}$ account for multiple scattering between the antennas and their environment. The influence of the backscattering from the receiving antenna~2 to antenna~1 through~$\mathbf{M}_{12}$, ref. Fig.~\ref{fig:SWF_Flow_graph_complete}, is neglected. Compared to \eqref{eq:S21_FS}, determining all necessary parameters of channel and antennas in \eqref{eq:S21_OB_org} numerically increases the computational effort significantly \cite{Pirkl2012}.

To make the SWF de-embedding approach for WBAN more straightforward and applicable with standard commercial simulation software, we will derive an alternative, single-step antenna de-embedding scheme in the following.

\begin{figure}
    \centering
    \includegraphics[width=0.95\columnwidth]{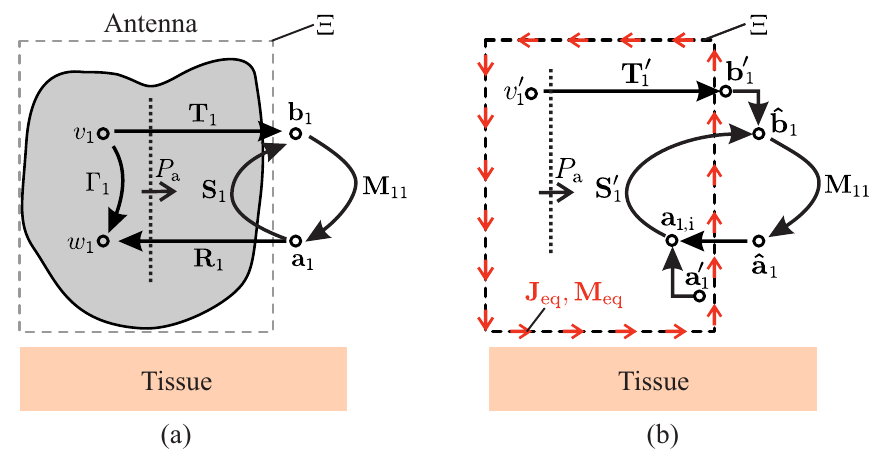}
    \caption{Antenna scattering matrix in WBAN scenario. (a) Original scenario; (b) Surface equivalence theorem: equivalent currents~$\mathbf{J_\mathrm{eq}}, \mathbf{M_\mathrm{eq}}$ on enclosure~$\Xi$ excite outgoing waves~$\mathbf{b_1^\prime}$ and incoming waves~$\mathbf{a_1^\prime}$}
    \label{fig:Flow_Graph_S11}
\end{figure}
{
The antenna de-embedding problem as depicted in Fig.~{\ref{fig:SWF_Flow_graph_complete}} can be reduced to the one shown in Fig.~{\ref{fig:Flow_Graph_S11}}(a) if the influence of the second antenna on the input reflection coefficient of the first antenna can be neglected.
The outgoing waves~$\mathbf{b}_1$ of the transmitting antenna are then determined by:
}
\begin{equation}
    \label{eq:b}
    \mathbf{b}_1=\mathbf{T}_1v_1+\mathbf{S}_1\mathbf{a}_1.
\end{equation}
{
If the radiated fields of an antenna are characterized trough $\mathbf{a}_1$ and $\mathbf{b}_1$, in {\eqref{eq:b}} there are two unknowns with $\mathbf{S}_1$ and $\mathbf{T}_1$. Thus, a direct solution as above in {\eqref{eq:T}} is not possible.
Therefore, the radiating antenna is virtually removed from the scenario and replaced by equivalent currents ($\mathbf{J}_{\mathrm{eq}}$,~$\mathbf{M}_{\mathrm{eq}}$) on an enclosing surface~$\Xi$, ref.~Fig.{\ref{fig:Flow_Graph_S11}}(b), according to the Huygens Equivalence Theorem\cite{Balanis2012}.
The interior of the surface~$\Xi$ is now assumed to be empty space, such as the antenna's SWF scattering matrix becomes equal to the unity matrix~$\mathbf{S}_1^\prime = \mathbf{I}$, because the origin reflects every incoming SWF as the corresponding outgoing SWF (similar as an open termination with guided waves).
}
Furthermore, the currents~$\mathbf{J}_{\mathrm{eq}}$ and~$\mathbf{M}_{\mathrm{eq}}$ are determined
with the constraint that they excite only outgoing waves ($\mathbf{a}_1^\prime=0$, ref.~Fig.~\ref{fig:Flow_Graph_S11}(b)).
If now the outgoing waves excited by~$\mathbf{J}_{\mathrm{eq}}$ and~$\mathbf{M}_{\mathrm{eq}}$ are chosen as \cite{Moerlein2022}:
\begin{equation}
\label{eq:swf_eqv}
    \mathbf{b}_1^\prime=\mathbf{b}_1-\mathbf{a}_1,
\end{equation}
the overall SWF coefficients~$\mathbf{\hat{b}}_1$ and~$\mathbf{\hat{a}}_1$ in the equivalent setup, ref. Fig.~\ref{fig:Flow_Graph_S11}(b), and thus the fields outside~$\Xi$ remain equal to the original setup in Fig~\ref{fig:Flow_Graph_S11}(a):\footnote{It is interesting to note here that although the excited currents radiate only outwards, the fields inside~$\Xi$ are non-zero and consist of the incoming waves~\mbox{$\mathbf{a_\mathrm{i}}=\mathbf{M}_{11}\mathbf{b}$}.}
\begin{equation}
    \mathbf{\hat{b}}_1=\mathbf{b}^\prime_1+\mathbf{I}\mathbf{a}_1=\mathbf{b}_1;\,\,\,\,\,\,\,\,\,\mathbf{\hat{a}}_1=\mathbf{M_{11}}\mathbf{\hat{b}}_1=\mathbf{a}_1.
\end{equation}
The computational complexity of \eqref{eq:b} is reduced, since $\mathbf{T}_1^\prime$ is now readily available from:
\begin{equation}
\label{eq:T_prime}
    \mathbf{b}_1^\prime=\mathbf{T}_1^\prime v_1^\prime.
\end{equation}
To keep the accepted port power $P_\mathrm{a}$ constant between the original domain, ref. Fig.~\ref{fig:Flow_Graph_S11}(a) and the equivalent problem, ref.~Fig. ~\ref{fig:Flow_Graph_S11}(b), the virtual port signal in \eqref{eq:T_prime} is defined as \mbox{$v_1^\prime=\sqrt{2P_\mathrm{a}}=\sqrt{|v_1|^2-|w_1|^2}$}. 
 
Instead of obtaining the coefficients~$a_j$ and~$b_j$ separately, the coefficients $b_j^\prime$ of the vector $\mathbf{b}_1^\prime$ in \eqref{eq:swf_eqv} can be calculated directly from the near field of the antenna using a regular wave~$\mathbf{F}_j^{(1)}$:
\begin{equation}
    \label{eq:b_prime}
    b_j^\prime = b_j - a_j = \frac{2k}{j\sqrt{\eta}} \left<\mathbf{E}, \mathbf{F}_j^{(1)*}\right>.
\end{equation}
Here, the SWF decomposition is performed on the surface~$\Xi$ with the notation:
\begin{equation}
    \label{eq_integral_definition}
    \left<\, \mathbf{u},\; \mathbf{v}\, \right> =
    \oiint_\Xi \; \left\{
          \mathbf{u} \times \left( \nabla \times \mathbf{v} \right)
        - \mathbf{v} \times \left( \nabla \times \mathbf{u} \right)
    \right\} \mathbf{\hat{n}} \; \mathrm{d}\Xi.
\end{equation}

Using regular SWF in the decomposition is also numerically superior compared to calculating~$\mathbf{b}$ or~$\mathbf{a}$ independently, as Santiago~et~al. have recently shown\cite{Santiago2019}. The receive vector~$\mathbf{R}_2^\prime$ of the equivalent problem can be found from~$\mathbf{T}_2^\prime$ using reciprocity  \cite{Hansen1988}. Finally, for adopting the channel response~$S_{21}$ in \eqref{eq:M} to the modified coefficients $\mathbf{b}^\prime$, an alternative mode-to-mode transmission matrix~$\mathbf{M}_{21}^\prime$ is defined as:
\begin{equation}
    \mathbf{a}^\prime_\mathrm{2}=\mathbf{M}_{21}^\prime(\mathbf{b}_\mathrm{1}-\mathbf{a}_\mathrm{1})=\mathbf{M}_{21}^\prime\mathbf{b}_\mathrm{1}^\prime,
\end{equation}
where~$\mathbf{a}^\prime_\mathrm{2}$ are the incoming waves at the location of the receiving antenna in the equivalent problem with removed antennas, ref. Fig.~\ref{fig:Flow_Graph_S11}(b).

By resolving the loop between $\mathbf{M}_{11}$ and $\mathbf{S}$ in Fig.~\ref{fig:Flow_Graph_S11}(b) according to signal flow graph theory, the relation to the original SWF channel transmission matrix is determined as:
\begin{equation}
    \mathbf{M}_{21}^\prime=\left(\mathbf{I}-\mathbf{M}_{22}\mathbf{S}_2^\prime\right)^{-1}\mathbf{M}_{21}\left(\mathbf{I}-\mathbf{S}_1^\prime\mathbf{M}_{11}\right)^{-1}.
\end{equation}
As can be seen from the equation, the adapted SWF channel transmission matrix~$\mathbf{M}_{21}^\prime$ already includes the influence of the backscattering of the channel for the equivalent problem with removed antennas.
The channel response of the equivalent problem is then calculated as:
\begin{equation}
\label{eq:S_21_OB_prime}
    S_{21}=\mathbf{R}_2^\prime\mathbf{M}_{21}^\prime\mathbf{T}_1^\prime.
\end{equation}
Concluding, \eqref{eq:S21_OB_org} has been rearranged to the simple form \eqref{eq:S_21_OB_prime} with only three parameters to be determined as in \eqref{eq:S21_FS}.

\section{Antenna Optimization Using SWF}
\label{sec:SWF_opt}
Up to this point, the SWF antenna de-embedding for WBAN is used solely as a modeling approach for enabling an efficient assessment of the behavior of different antennas in a channel. In contrast to the concept of using DD channel models and optimizing antennas by means of their on-body gain, antenna optimization with SWF modeling still is only trial and error. 
However, as will be shown in the following, SWF modeling can also be used for antenna optimization.

To optimize the transmitting antenna, the goal is to maximize the power of the incoming waves~$\mathbf{a}_2^\prime$ at the receiver position. Without loss of generality, the Euclidean norm of the outgoing wave coefficient vector can be set to ~$\|\mathbf{b}_1^\prime\|_2=1$ and the optimization problem can be formulated as:
\begin{equation}
\label{eq:opt_problem}
    \mathrm{max}\,\left\|\mathbf{a}_2^\prime\right\|_2=\operatorname*{max}_{\|\mathbf{b}_1^\prime\|_2=1}\,\left\|\mathbf{M}_{21}^\prime \mathbf{b}_1^\prime\right\|_2.
\end{equation}
Assuming a known SWF channel transmission matrix~$\mathbf{M}_{21}^\prime$, the optimal excitation~$\mathbf{b}^\prime_{1,\mathrm{opt}}$ is sought.
By using the definition of the spectral norm, the optimum can be calculated analytically as \cite{Kiltz2004}:
\begin{equation}
    \operatorname*{max}_{\left\|\mathbf{b}_1^\prime\right\|_2=1} \left\|\mathbf{M}^\prime_{21}\mathbf{b}_1^\prime\right\|_2 = \sqrt{\left\|\mathbf{M}^{\prime \mathrm{H}}_{21}\,\mathbf{M}^\prime_{21}\right\|_2} = \sqrt{\lambda_\mathrm{max}},
\end{equation}
where~$\lambda_\mathrm{max}$ is the largest eigenvalue of~$\mathbf{M}_{21}^{\prime \mathrm{H}}\,\mathbf{M}_{21}^\prime$. Finally, the corresponding eigenvector is the optimal excitation~$\mathbf{b}^\prime_{1,\mathrm{opt}}$.
For calculating the transmission coefficient \eqref{eq:S_21_OB_prime} with the optimal excitation,~$\mathbf{T}^\prime_{1,\mathrm{opt}}$ is calculated using \eqref{eq:T_prime} as:
\begin{equation}
    \label{eq:T_opt}
    \mathbf{T}^\prime_{1,\mathrm{opt}} = \frac{1}{\sqrt{2P_\mathrm{a}}} \mathbf{b}^\prime_{1,\mathrm{opt}}.
\end{equation}
If the optimal antenna is assumed to be lossless, the accepted power $P_\mathrm{a}$ is identical to the radiated power $P_\mathrm{rad}$ going through~$\Xi$ in Fig.~\ref{fig:Flow_Graph_S11}(b) and can be calculated :
\begin{equation}
\label{eq:P_rad1}
    P_\mathrm{a}=P_{\textrm{rad}} = \frac{1}{2} \left(\|{\hat{\mathbf{b}_1}}\|_2^2 - \|{\hat{\mathbf{a}_1}}\|_2^2\right)\,.
\end{equation}
Using the signal flow graph in Fig.~\ref{fig:Flow_Graph_S11}(b), this can be written as:
\begin{equation}
\label{eq:P_rad2}
 P_\mathrm{a} = \frac{1}{2} + \operatorname{Re}\left\{ \left(\left( \mathbf{I}-\mathbf{M}_{11} \right)^{-1} \mathbf{M}_{11} \mathbf{b}_1^\prime \right)^{\textrm{H}} \mathbf{b}_1^\prime \right\} \:.
\end{equation}

While the vector~$\mathbf{T}^\prime_\mathrm{1,opt}$ already represents the optimal solution (in a spherical basis), antenna designers typically do not think in terms of spherical wave coefficients, so $\mathbf{T}^\prime_\mathrm{1,opt}$ is a quite abstract parameter. Therefore, a visualization of the calculated optimal antenna properties for a given channel will be derived. For this matter, the on-body gain as defined in \cite{Berkelmann2021} is calculated for the optimal antenna.
Hence, for each individual spherical mode~$j$ to be considered, equivalent currents $(\mathbf{J}_j, \mathbf{M}_j)$ on surface~$\Xi$, ref.~Fig.\ref{fig:Flow_Graph_S11}(b), are defined analytically:
\begin{equation}
\label{eq:opt_eq_currents_J}
    \mathbf{J}_j=\mathbf{n}\times\left(\frac{ikv}{\sqrt{\eta}}\mathbf{F}^{(4)}_{\overline{j}}(\mathbf{r}_\Xi)\right),
\end{equation}
\begin{equation}
\label{eq:opt_eq_currents_M}
    \mathbf{M}_j=-\mathbf{n}\times\left(k\sqrt{\eta}v\mathbf{F}^{(4)}_{j}(\mathbf{r}_\Xi)\right),
\end{equation}
with~$i$ being the imaginary unit and $\mathbf{r}_\Xi$ the position vectors on the surface~$\Xi$.
The on-body far fields radiated by the optimal antenna can then be calculated by:
\begin{IEEEeqnarray}{rCl}
    \mathbf{E}&=&\oiint\limits_{\Xi} \sum_{j=1}^J b^\prime_{j,\mathrm{opt}}\left( \overline{\mathbf{G}}_\mathrm{J}\cdot\mathbf{J}_j
    +\overline{\mathbf{G}}_\mathrm{M}\cdot\mathbf{M}_j\right)\mathrm{d}\Xi\IEEEeqnarraynumspace
\end{IEEEeqnarray}
where~$\overline{\mathbf{G}}_\mathrm{J}$ and~$\overline{\mathbf{G}}_\mathrm{M}$
denote the dyadic Green's functions for the on-body case approximated by a tissue half-space as derived in \cite{Berkelmann2021}, and~$b^\prime_{j,\mathrm{opt}}$ the individual coefficients in~$\mathbf{b}^\prime_\mathrm{opt}$.
The on-body gain is then calculated as \cite{Berkelmann2021}:
\begin{equation}
\label{eq:onbody_gain}
     G_{\mathrm{B}}=\frac{\pi\rho^{2} E_\perp^{2}}{\eta_0  |F(\rho)|^2 P_\mathrm{a}},
\end{equation}
where~$\rho$ is the radial distance from the antenna parallel to the assumed tissue halfspace at which the on-body far fields are calculated,~$E_\perp=\mathbf{n}_\mathrm{t}\cdot\mathbf{E}$ the normal component of the \mbox{E-field} with respect to the body tissue and~$F(\rho)$ the Sommerfeld attenuation factor \cite{Berkelmann2021} for normalizing the range dependent losses due to the tissue.
{As will be seen in the following, this quantity can be analyzed equivalently to the free space antenna radiation pattern in the form of a polar diagram and thus provide important information about the directional properties of the antennas for the on-body channel.}

\section{Example Application}
\label{sec:application}
\begin{figure}
    \centering
    \includegraphics[width=0.95\columnwidth]{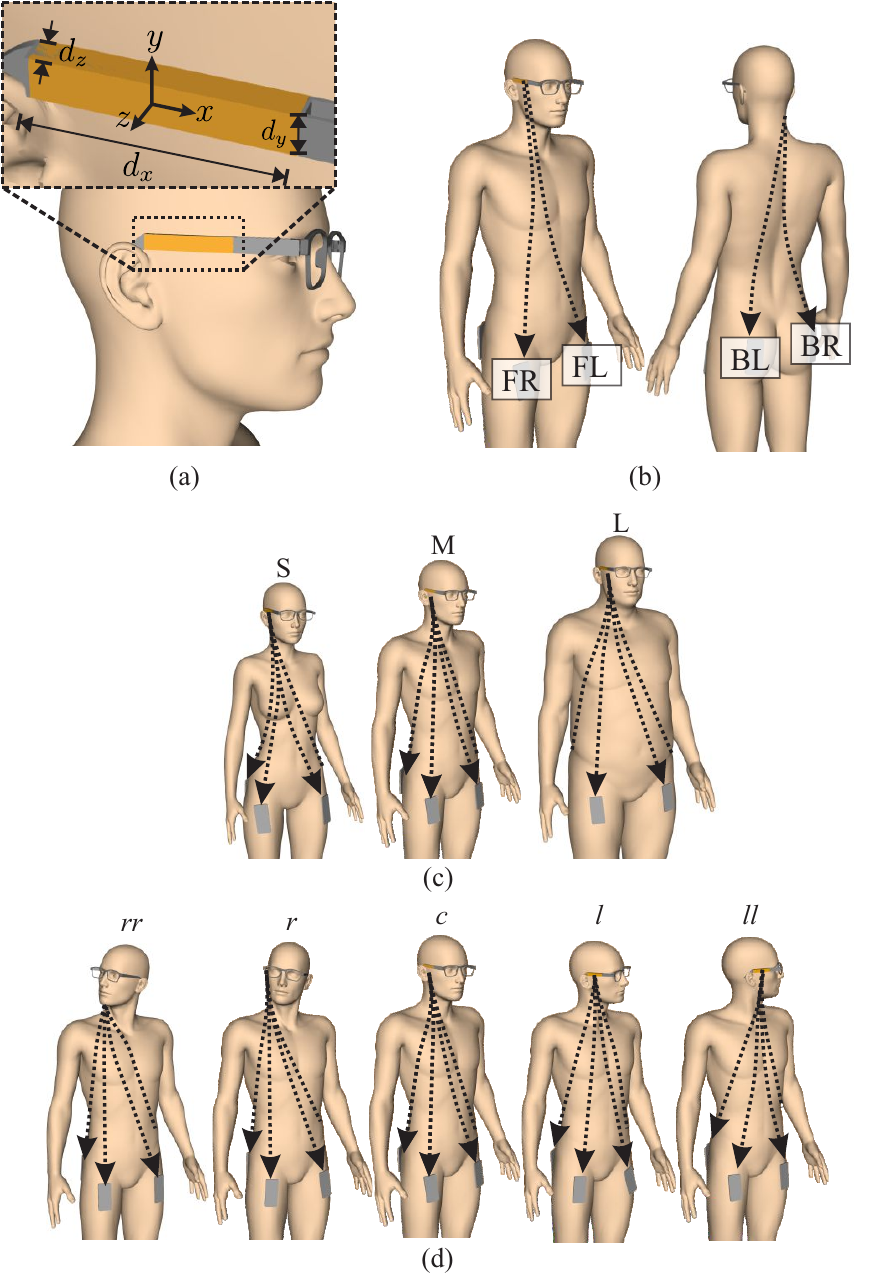}
    \caption{Example application of antenna for eye-wear application: (a) Antenna integration volume; (b) Receiver locations: front-left~(FL) to back-right~(BR); (c) Anatomies: small~(S) to large~(L); (d) Body poses (head rotation): right-right~(\textit{rr}) to left-left~(\textit{ll})}
    \label{fig:example_app_anatomies_poses}
\end{figure}
To implement and evaluate the methods as derived above, an antenna for eye-wear applications in the 2.4\,GHz ISM-band is designed and tested as an example in the following.
The antenna is supposed to be integrated in the right glasses temple as can be seen in Fig.~\ref{fig:example_app_anatomies_poses}(a). Thus, the available antenna integration space has a \mbox{volume of $(d_x\times d_y\times d_z)=(60 \times 10 \times 7.5)\,\mathrm{mm}^3$,} respectively~$(0.5 \times 0.08 \times 0.06)\,\lambda^3$ at~$f=2.45$\,GHz. 

In the following, the node with the antenna integrated into the glasses is referred to as the transmitter. The RF link from the glasses to a receiver (e.g. a smartphone), which is assumed to be carried in a trouser pocket, is optimized.
Given the selected application, the antenna optimization is performed exclusively for the transmitter. As the receiver, a top-loaded monopole antenna (isotropic on-body gain \mbox{$G_\mathrm{B}=1.7$\,dB}, not shown) is utilized as a simplified placeholder, e.g. of a smartphone. Four different possible positions of the receiver are considered, ref.~Fig.~\ref{fig:example_app_anatomies_poses}(b). The modeling approach should take the influence of posing and different anatomies into account. Therefore, considering different anatomies, three different body phantoms are chosen as depicted in Fig.~\ref{fig:example_app_anatomies_poses}(c), an average tall woman ($h=1.6$\,m) and man ($h=1.8$\,m) were considered, as well as a particularly tall, portly man ($h=2$\,m).
As can be seen in Fig.~\ref{fig:example_app_anatomies_poses}(d), with regard to posing, five different head rotation angles are considered, since this parameter is assumed to have the biggest effect on the on-body propagation in the chosen example. Thus, in total 60 different channel matrices are considered.

\subsection{SWF Channel Modeling}
\begin{figure}
    \centering
    \includegraphics[width=0.95\columnwidth]{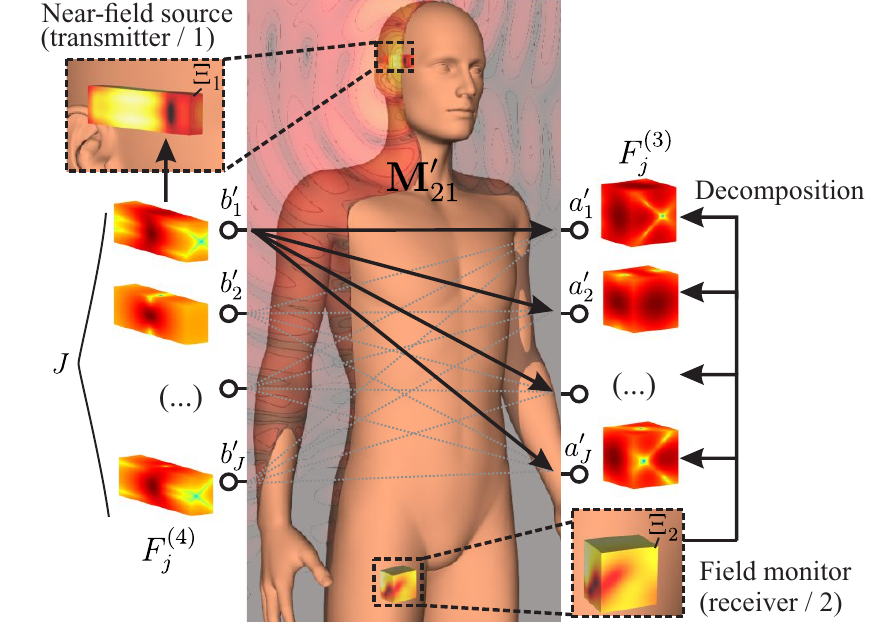}
    \caption{SWF channel modeling: A near-field source is used to excite the SWF modes $b_j$ sequentially (top left). At the receiver position (bottom right), the incident fields are recorded and decomposed into vectors of incoming~waves~$\mathbf{a}$}
    \label{fig:swf_channel_modeling}
\end{figure}

For the channel modeling with regard to antenna \mbox{de-embedding} as derived in Section~\ref{sec:SWF_WBAN}, the transmission matrix~$\mathbf{M}^\prime_{21}$ needs to be calculated. 
The numerical calculation procedure is realized as a sequential excitation of each SWF mode in place of the antenna by a near-field source, which is usually part of standard EM modeling software. Our implementation is outlined in Fig.~\ref{fig:swf_channel_modeling}.
Here, the near-field source must be transparent and non-scattering so that the incoming waves can pass through unhindered. This is necessary so that the assumptions made for the equivalent problem, ref.~Fig~\ref{fig:Flow_Graph_S11}(b), are valid.
The surface~$\Xi_1$, that encloses the transmitting antenna, is used for the excitation, ref. Fig.~\ref{fig:swf_channel_modeling}~(top left). Based on \eqref{eq:e_total_sum}, for each mode~$j$, the fields to excite on~$\Xi_1$ are calculated as:
\begin{equation}
    \mathbf{E}_j=k\sqrt{\eta}\mathbf{F}_j^{(4)},
\end{equation}
where compared to \eqref{eq:e_total_sum} a single entry $j$ is set as $b_j=1$ and all other entries in $\mathbf{b}_1$ are zero.
The radiated fields are calculated numerically for each excited mode individually. Thus, in total, $J$~simulations need to be performed. In our implementation, this is done using the FDTD solver of EMPIRE XPU \cite{empire}. In the simulations, the fields on the enclosing surfaces~$\Xi_2$ around each receiving antenna location are recorded, ref.~Fig.~\ref{fig:swf_channel_modeling} (bottom right). 
These fields on~$\Xi_2$ represent the incoming waves at the receiver and thus are decomposed into vectors of incoming wave coefficients~$\mathbf{a}_2^\prime$. Because a single mode~$b_j^\prime$ is excited in each simulation, each vector~$\mathbf{a}_2^\prime$ directly forms a column of the channel matrix~$\mathbf{M}_{21}^\prime$. In Fig.~\ref{fig:swf_channel_modeling} (center), this is outlined by the arrows connecting a single excited outgoing spherical wave~$b^\prime_j$ with each incoming mode~$a^\prime_j$ at the receiver.

\subsection{Antenna Optimization}
\label{sec:opt_antenna_swf}
\begin{figure}
    \centering
    \includegraphics[width=0.7\columnwidth]{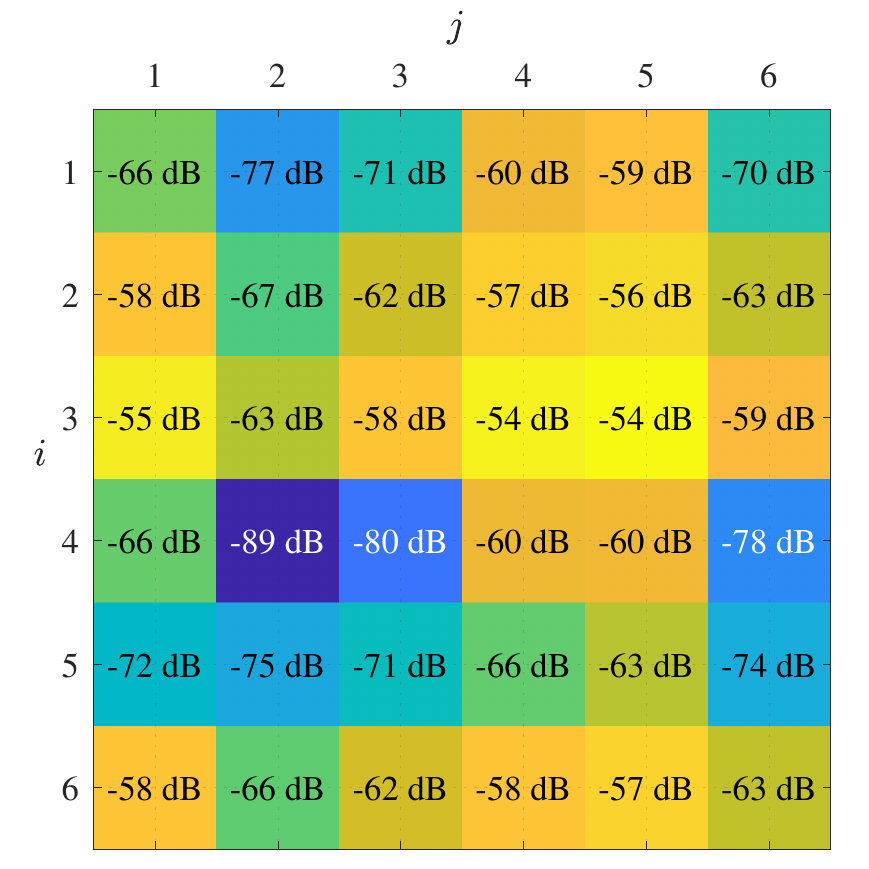}
    \caption{Magnitudes of the entries of a transmission matrix~$\mathbf{M}_{21}^\prime$ (example: receiver FL with phantom M/c, ref.~Fig.~\ref{fig:example_app_anatomies_poses}). The columns~$j$ correspond to the excitation of~$\mathbf{b}^\prime$ at the transmitter, rows~$i$ to the incoming waves~$\mathbf{a}^\prime$ at the receiver}
    \label{fig:M_21_example}
\end{figure}

To find optimal antennas as discussed in Section~\ref{sec:SWF_opt}, additionally the channel's spherical mode reflection property in form of~$\mathbf{M}_{11}$ in \eqref{eq:P_rad2} needs to be determined. For this purpose, the total fields on the surface~$\Xi_1$ enclosing the antenna resulting from the simulation are also recorded when the individual modes~$b^\prime_j$ are excited and decomposed into~$\mathbf{\hat{b}}_1$. Assembling all column vectors~$\mathbf{\hat{b}}_1$ in the matrix~$\mathbf{\hat{B}}_1$, the channel reflection matrix can be calculated as:
\begin{equation}
    \mathbf{M}_{11}=\mathbf{I}-\hat{\mathbf{B}}_1^{-1}.
\end{equation}
Due to the small antenna integration volume with the largest dimension of about~$d=0.5\lambda$ in the example application, ref.~Fig.~\ref{fig:example_app_anatomies_poses}, it can be assumed that the antenna is mainly represented by the first six spherical wave modes (corresponding to small electric and magnetic dipoles).
Hence, the channel matrices~$\mathbf{M}_{21}^\prime$ for the antenna optimization are also calculated for the first six spherical wave modes only. The utilized near-field source has an edge length of 16\,mm, with its center coinciding with the center of the antenna integration space. In Fig.~\ref{fig:M_21_example}, one of the calculated channel matrices is shown as an example. For antenna optimization, the columns with the greatest norm (the greatest power at the receiver) are now searched for. At this point, it is not yet possible to define clear antenna design goals (e.g. for exciting a selected mode), as obviously multiple solutions exist and multiple modes are to be combined eventually.
However, based on the channel matrices $\mathbf{M}_{21}^\prime$, optimal properties for the transmitting antenna can be calculated using the optimization method derived in Section~\ref{sec:SWF_opt}.

In general, the optimal solution $\mathbf{b}_{\mathrm{opt}}$ may consist of an arbitrary combination of excited spherical waves. To realize such a general antenna, both TE and TM modes would have to be excited simultaneously by that antenna. This requires so-called Huygens antennas, whose design is very challenging~\cite{Tang2016}. Therefore, we restrict the solution space and perform the optimization individually for either only TE or TM Modes. 
As will be shown in the following, this also ensures better comparability with the practical design evaluated.

\begin{figure}
    \centering
    \includegraphics[width=0.9\columnwidth]{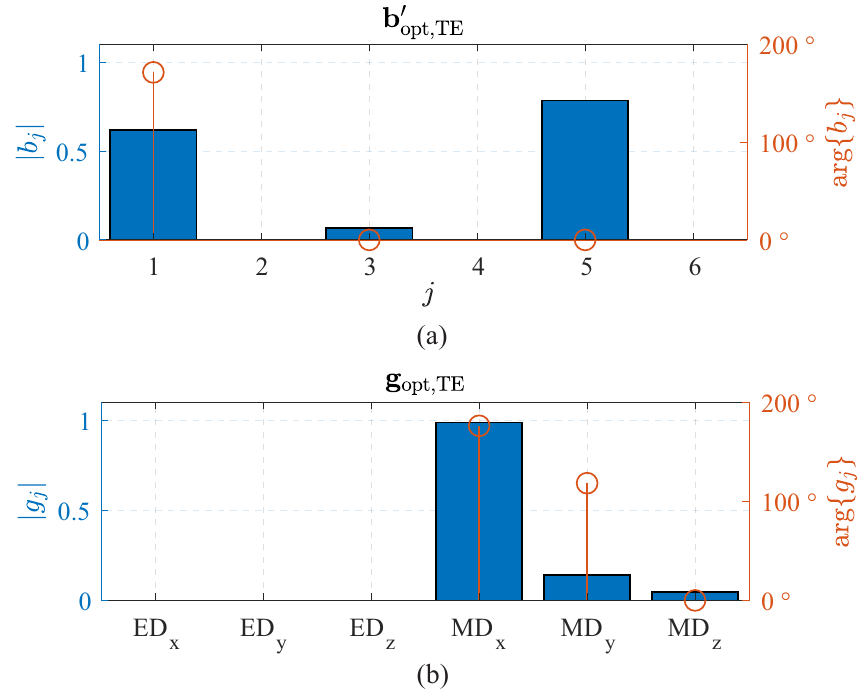}
    \caption{Optimized spherical wave excitation for the example application, ref.~Fig.\ref{fig:example_app_anatomies_poses} (weighted superposition, ref.~\eqref{eq:superpos_opt}): (a) optimal TE-mode excitation; (b) associated weighting coefficients for small magnetic (MD) and electric (ED) dipoles}
    \label{fig:optimal_mean_TE_coef}
\end{figure}

The optimal spherical wave excitation $\mathbf{b}^\prime_{\textrm{opt}}$ is calculated individually for all channel matrices $\mathbf{M}_{21}^\prime$. A global optimum $\mathbf{b}_{\mathrm{opt,g}}^\prime$ can then be defined by the superposition of the individual optimal antenna properties:
\begin{equation}
    \label{eq:superpos_opt}
    \mathbf{b}_{\mathrm{opt,g}}^\prime = \sum_{k=1}^{60} p_k \mathbf{b}_{\mathrm{opt},k}^\prime,
\end{equation}
whereby $k$ is the index describing the scenarios, ref. Fig.~\ref{fig:example_app_anatomies_poses}, $p_k$ is the weighting (e.g. based on the likelihood) of the $k^\mathrm{th}$ scenario and $\mathbf{b}_{\mathrm{opt},k}$ is the optimum in the individual scenario. In this case, we assume that all scenarios are equally weighted $p_k = 1/60$.
In the chosen example, the optimum calculated for TE-mode excitation gives a higher power of incoming waves at the receivers on average. Therefore, only this optimum is considered in the following. Fig.~\ref{fig:optimal_mean_TE_coef}(a) depicts the global optimum for TE-mode excitation $\mathbf{b}^\prime_\mathrm{opt,TE}$ which was calculated using \eqref{eq:superpos_opt}. From the first six SWF as depicted in Fig.~\ref{fig:optimal_mean_TE_coef}(a), the weighting coefficients~$\mathbf{g}_\mathrm{opt,TE}$ for equivalent dipoles can also be obtained \cite{Hansen1988} as depicted in in Fig.~\ref{fig:optimal_mean_TE_coef}(b). As can be seen, for the chosen example, the solution of the optimization using TE modes only can be approximated very well by a single $x$-directed magnetic dipole, which can be used subsequently as an antenna design guideline. However, the design guideline for the optimal antenna cannot always be expected to be as clear as in the chosen example. In those cases, the on-body gain pattern for the optimal excitation can be calculated using \eqref{eq:opt_eq_currents_J} - \eqref{eq:onbody_gain}.
\subsection{Antenna Design}
\begin{figure}
    \centering
    \includegraphics[width=0.95\columnwidth]{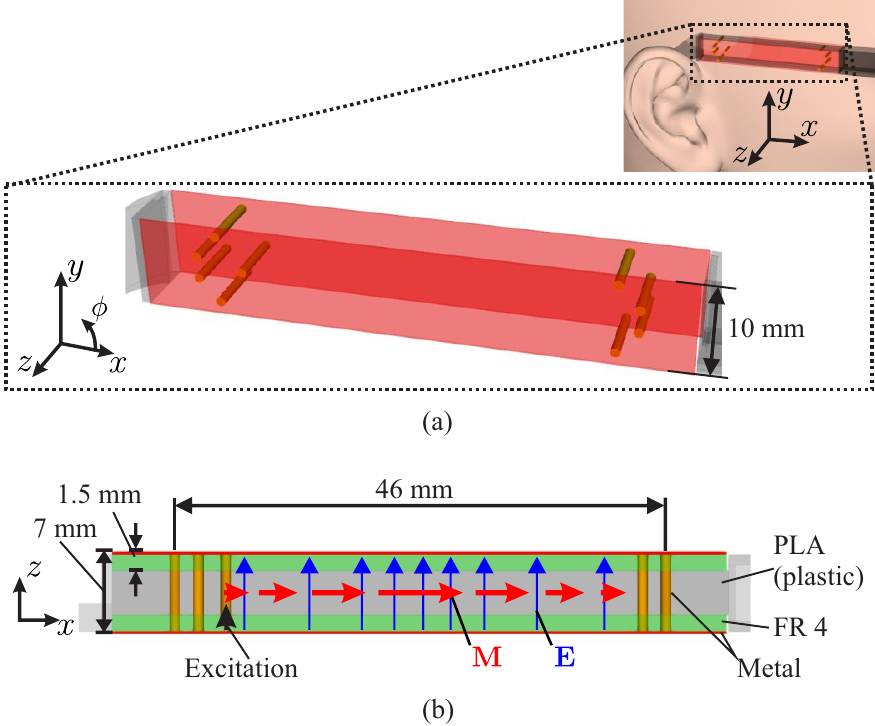}
    \caption{Magnetic dipole antenna design for example eyewear application: (a) Antenna structure consisting of two parallel plates which are shorted at the ends through multiple pins; (b) $xz$-cut of the designed antenna with excited $\mathbf{E}$-fields and equivalent magnetic currents $\mathbf{M}$.}
    \label{fig:magnetic_dipole_antenna}
\end{figure}

\begin{figure}
    \centering
    \includegraphics[width=0.9\columnwidth]{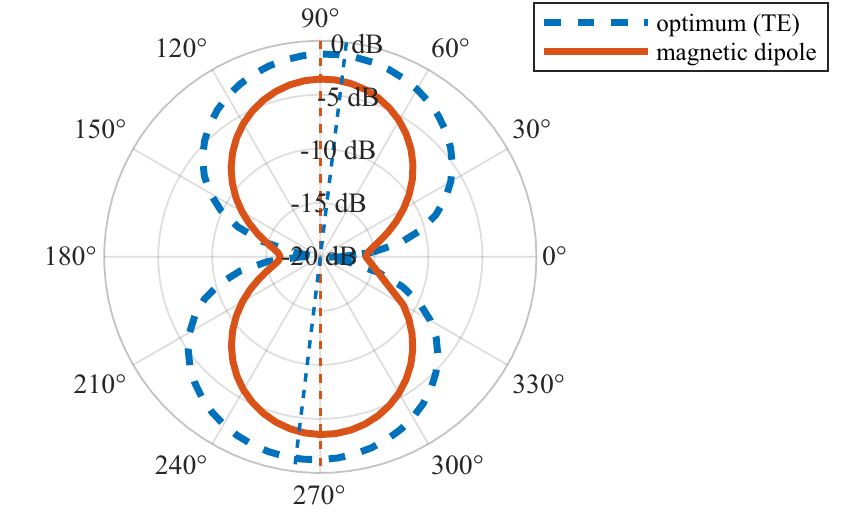}
    \caption{On-body gain \cite{Berkelmann2021} ($xy$-plane) of the optimal TE mode excitation~(ref.~Fig.~\ref{fig:optimal_mean_TE_coef}), compared to magnetic dipole design~(ref.~Fig.~\ref{fig:magnetic_dipole_antenna}). The dashed lines mark the axes with maximum gain.}
    \label{fig:onbody_gain_magnetic_dipole_vs_TE_opt}
\end{figure}

Based on the calculated optimal excitation for the example application, the antenna can be designed. An antenna that realizes an $x$-directed magnetic current required for the optimal antenna, is depicted in Fig.~\ref{fig:magnetic_dipole_antenna}(a). A similar design is described in detail in \cite{Berkelmann2019}. It consists of two parallel metal plates (realized on an FR4 substrate) with a length of approximately $0.5\lambda$, which are shorted through multiple pins at both ends. The excited $\mathbf{E}$-field distribution with its maximum at the center of the antenna in $x$-direction is outlined in the $xz$-cut of the antenna structure in Fig.~\ref{fig:magnetic_dipole_antenna}(b).
The radiated fields can equivalently be represented by an $x$-directed magnetic half-wave dipole of magnetic currents $\mathbf{M}$ which are also outlined in Fig.~\ref{fig:magnetic_dipole_antenna}(b).
The relationship to the optimal antenna is particularly evident in the on-body gain pattern as depicted in Fig.~\ref{fig:onbody_gain_magnetic_dipole_vs_TE_opt}. For the magnetic dipole antenna structure, the on-body gain is calculated directly from the simulated antenna's near field as described in \cite{Berkelmann2021}. The on-body gain for the optimal antenna is calculated by \eqref{eq:opt_eq_currents_J}-\eqref{eq:onbody_gain}.
As can be seen, the chosen magnetic dipole antenna has an on-body gain pattern very similar to the optimum case. However, due to losses of the antenna structure, the gain is reduced by approximately~2\,dB compared to the theoretical optimum.

\subsection{Antenna Channel Embedding}
\begin{figure}
    \centering
    \includegraphics[width=0.9\columnwidth]{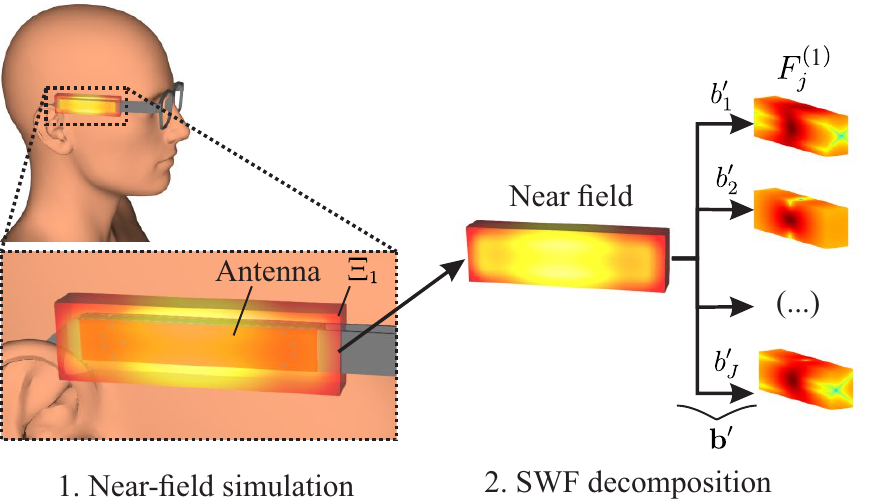}
    \caption{SWF antenna modeling: The near field of the antenna is first extracted from a simulation and the fields are then decomposed using \eqref{eq:b_prime}}
    \label{fig:SWF_antenna_decomposition}
\end{figure} 
To evaluate the behavior of the chosen antenna design inside the different SWF channel models, first, the antenna's near field is extracted from a simulation as depicted in Fig.~\ref{fig:SWF_antenna_decomposition}. Secondly, the antenna's radiated fields are decomposed into $\mathbf{b}^\prime$. 
In the evaluated example, the field monitor $\Xi$ surrounding the antenna, respectively the near-field source in the channel simulation, has an edge length of 62\,mm. To find a sufficient truncation for the number of modes $J$ utilized for modeling the designed antenna, the convergence of the decomposition can be checked based on the norm $\|\mathbf{b}^\prime\|_2$ for different truncation numbers.

\begin{figure}
    \centering
    \includegraphics[width=0.9\columnwidth]{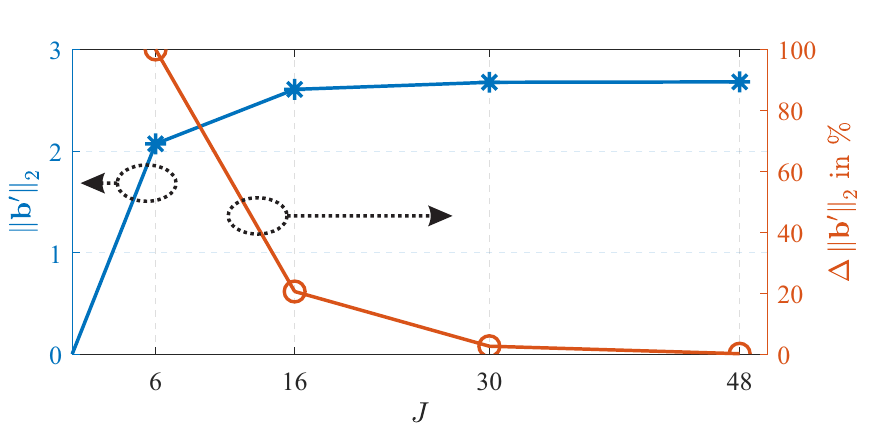}
    \caption{Convergence of the SWF antenna decomposition in dependence of mode truncation of the magnetic dipole antenna}
    \label{fig:mode_truncation}
\end{figure}

\begin{figure}
    \centering
    \includegraphics[width=0.9\columnwidth]{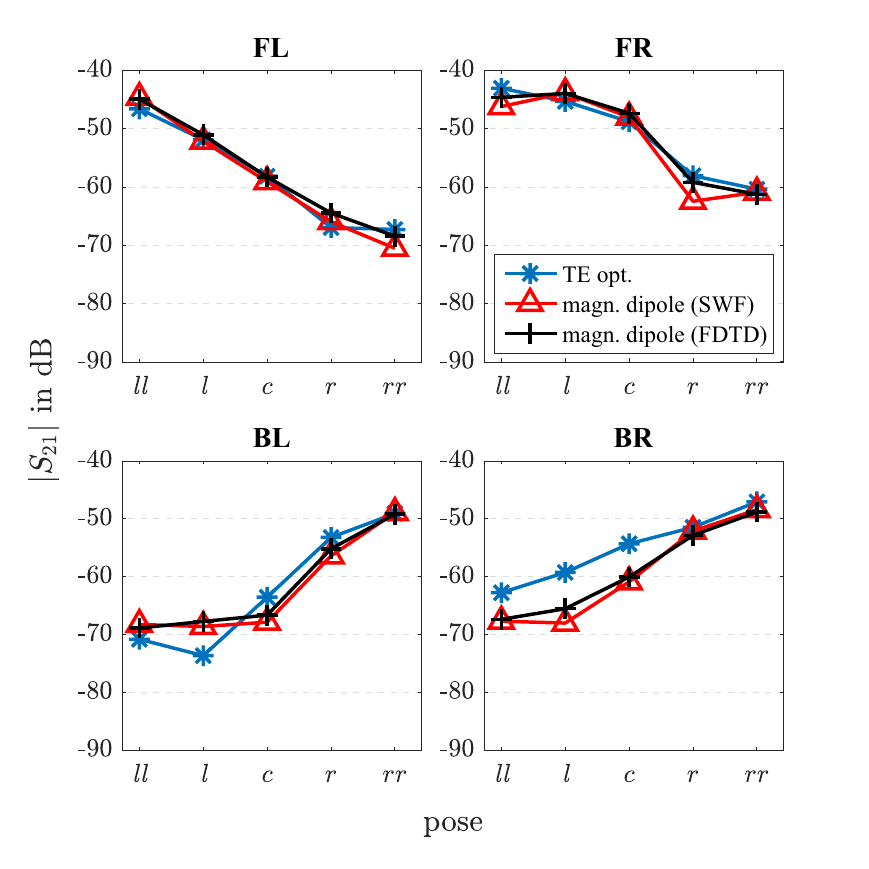}
    \caption{Transmission coefficient $|S_{21}|$ (average of different anatomies) to different receivers (FL-BR) in dependency of the body pose (\textit{ll}-\textit{rr}) with magnetic dipole antenna compared to theoretical optimal TE-mode antenna}
    \label{fig:swf_vs_fdtd}
\end{figure}

As depicted in Fig.~\ref{fig:mode_truncation}, the power of the excitation $\|\mathbf{b}^\prime\|_2$ converges against a limit with increasing $J$. Since the exact value of this limit is not known, the SWF decomposition is performed step-wise with an increasing mode truncation number~$J$.\footnote{It makes sense to use complete sets of a certain grade of SWF, e.g. $J=6$ for dipole modes only and $J=16$ to additionally include all quadrupole modes, etc., ref.~\cite{Hansen1988}} With each step $n$ for increasing $J$ the relative difference of $\|\mathbf{b}^\prime\|_2$ with the previous step ($n-1$) is evaluated:
\begin{equation}
    \Delta \|\mathbf{b}^\prime\|_2=\frac{\|\mathbf{b}_n^\prime\|_2-\|\mathbf{b}_{n-1}^\prime\|_2}{\|\mathbf{b}_n^\prime\|_2}.
\end{equation}
Thus, this quantity approaches zero when convergence is achieved. Evaluating the convergence with the designed magnetic dipole antenna in Fig.~\ref{fig:mode_truncation}, it can be concluded that as expected the antenna mainly excites SWF modes up to $J=6$. However, at the next higher number of $J=16$ the difference is still about $\Delta\|\mathbf{b}^\prime\|_2\approx20\,\%$. A possible reason for this is that the phase center of the antenna is not congruent with the origin of the SWF decomposition~\cite{Hansen1988}. Therefore, $J=30$ is used for the embedding of the antennas in the following.
Finally, the antenna's transmission vector $\mathbf{T}^\prime$ is calculated from $\mathbf{b}^\prime$ according to \eqref{eq:T_prime} by normalizing it to the accepted power $P_\mathrm{a}$, is readily available from the numerical simulation of the antenna conducted for calculating the near field as depicted in Fig.~\ref{fig:SWF_antenna_decomposition}. The monopole antenna, which is used at the receiver side, was characterized equivalently to calculate $\mathbf{R}^\prime$. As mentioned before, the receiving antenna (e.g. integrated inside a smartphone) cannot be adjusted concerning the chosen example application and is thus not evaluated in the analysis. Now that all parameters in \eqref{eq:S_21_OB_prime} are known, the transmission coefficient $S_{21}$ can be calculated with the designed antenna embedded into the different channels to evaluate the antenna's performance with regard to posing.
In Fig.~\ref{fig:swf_vs_fdtd}, the results calculated for the antenna embedded into all considered channels are visualized. The results of the magnitude of the transmission coefficient to the four receiver locations (FL-BR, ref.~Fig.~\ref{fig:example_app_anatomies_poses}) are depicted in separate plots. Each of the four plots in Fig.~\ref{fig:swf_vs_fdtd} depicts the transmission coefficient to one of the receivers in dependency of the body pose ($ll$-$rr$, ref.~Fig~\ref{fig:example_app_anatomies_poses}). With regard to varying anatomy, the average of the results with the three different sized phantoms (ref.~Fig.~\ref{fig:example_app_anatomies_poses}) is considered in Fig.~\ref{fig:swf_vs_fdtd}. The illustrated results reveal that the average transmission coefficient varies by up to 30\,dB due to different body poses considered in the evaluation. At the receivers located in front of the torso (FL, FR), the highest magnitudes of the transmission coefficient are observed in the pose $ll$ with the head turned all the way to the left so that the antenna integrated into the right glasses temple is facing to the front of the torso. Conversely, in the pose $rr$, the transmitter and receiver are on different sides of the body, hence their transmission coefficients are the lowest. With the receivers on the backside (BL, BR), this behavior is reversed.

To validate the results obtained from the SWF modeling, in Fig.~\ref{fig:swf_vs_fdtd}, the transmission coefficient obtained from a conventional FDTD simulation of the whole system for each scenario is depicted as well. As can be seen, only marginal deviations can be observed between the FDTD results and the SWF modeling approach. Additionally, the results calculated for the optimal TE mode excitation, ref. Sec.~\ref{sec:opt_antenna_swf}, are depicted. The very similar slopes of the pose-dependent transmission coefficients show the close relationship of the designed magnetic dipole antenna to the optimal TE mode excitation. Only at the receiver position BR, significant deviations of the slope with the designed antenna compared to the theoretical optimum can be observed. This can be attributed to the fact that the radiation pattern of the optimum antenna is slightly tilted compared to the straight vertical orientation of the designed magnetic dipole antenna, ref.~Fig~\ref{fig:onbody_gain_magnetic_dipole_vs_TE_opt}. The influence of this tilt becomes more obvious if the optimal radiation pattern is explicitly calculated for the link with  receiver BR (tilt of about $30^\circ$, not shown), rather than the overall optimum as depicted in Fig.~\ref{fig:onbody_gain_magnetic_dipole_vs_TE_opt}.

\section{Antenna Performance Evaluation}
\label{sec:measurements}
\begin{figure}
    \centering
    \includegraphics[width=0.9\columnwidth]{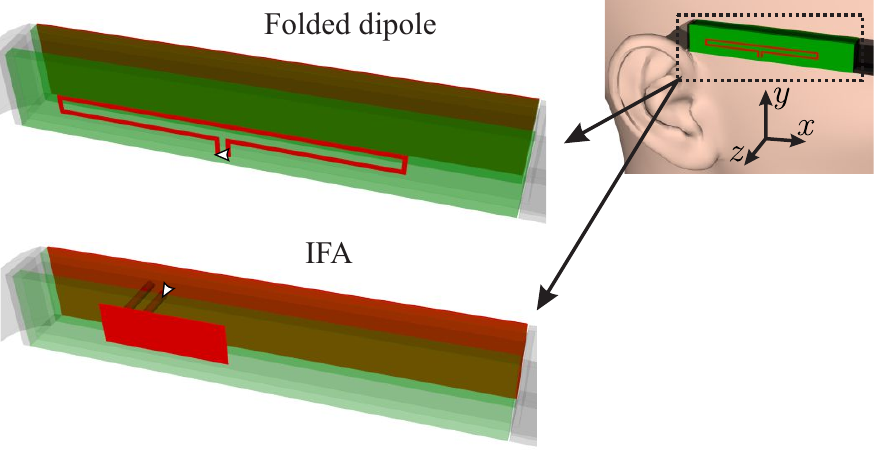}
    \caption{Additional antenna designs for evaluation of the SWF modeling method}
    \label{fig:add_antennas}
\end{figure}

{In the following, on the one hand, the implemented approach of SWF antennas and channel modeling for the selected example is further validated in comparison to a measurement campaign. On the other hand, it is shown how the system performance with different antennas can be compared by the possibility of the SWF methods of efficiently considering many different channel scenarios.}
Therefore, two additional antenna designs are evaluated for the eye-wear example application as described above. In addition to the designed magnetic dipole antenna (ref.~Fig.~\ref{fig:magnetic_dipole_antenna}) based on the SWF antenna optimization, a standard folded half-wave dipole antenna as well as an inverted-F antenna (IFA) as depicted in Fig.~\ref{fig:add_antennas} are utilized.

\begin{figure}
    \centering
    \includegraphics[width=0.99\columnwidth]{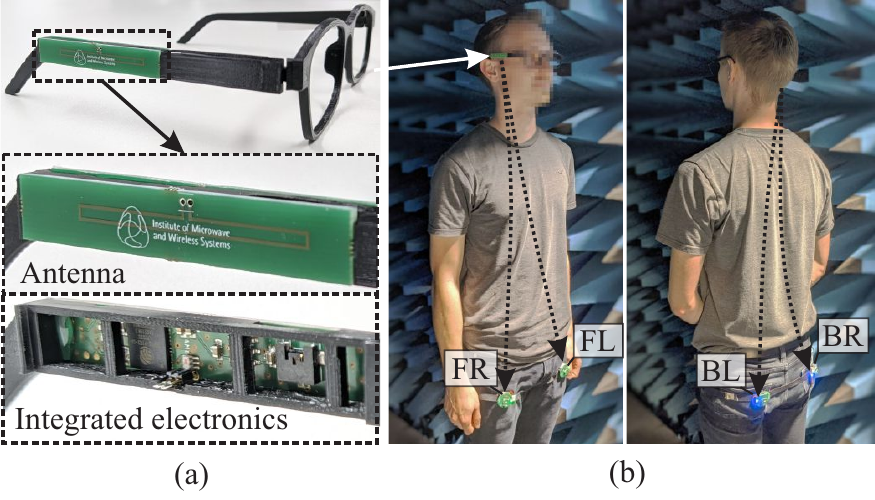}
    \caption{(a) Physical prototype of the evaluated antennas; (b) Measurements with human test subjects inside an anechoic chamber}
    \label{fig:prototype_picture}
\end{figure}
\subsection{{Measurement Setup}}
Physical prototypes of all three antennas were realized as depicted in Fig.~\ref{fig:prototype_picture}(a) and integrated along with a battery-powered Bluetooth transmitter circuit (Espressif ESP32-PICO-D4) into the glasses temples. Likewise, the four receivers were equipped with active Bluetooth transceivers. The transmission coefficients are estimated over the air (OTA) with the Bluetooth received signal strength indicator (RSSI). This way, cable effects are avoided. Measurements were performed with five human test subjects (all male, \mbox{body heights: \mbox{$h\in[172; 176; 178; 189; 192]$\,cm})} inside an anechoic chamber as depicted in Fig.~\ref{fig:prototype_picture}(b). With each test subject, multiple measurements in the five different body poses as defined in Fig.~\ref{fig:example_app_anatomies_poses} were taken.
To compensate for uncertainties in the measurements with regard to the transmitting power of the transceivers and possible antenna mismatching, a calibration factor is introduced for each antenna. It is calculated as the average magnitude of the transmission coefficient obtained from the numerical simulations divided by the average of all measured RSSI values.
As illustrated in Fig.~\ref{fig:example_meas_boxplot_vs_means}, the measured and calibrated transmission coefficients for each scenario were evaluated statistically. The average transmission coefficient (depicted as bold markers) is calculated depending on the pose at each receiver by combining the results of all five test subjects. The results in Fig.~\ref{fig:example_meas_boxplot_vs_means} also reveal that presumably due to fading, the observed transmission coefficient in individual measurements is sometimes 10\,-\,20\,dB below the calculated average.

\subsection{{KPI: Connection Loss Probability}}
\begin{figure}
    \centering
    \includegraphics[width=0.9\columnwidth]{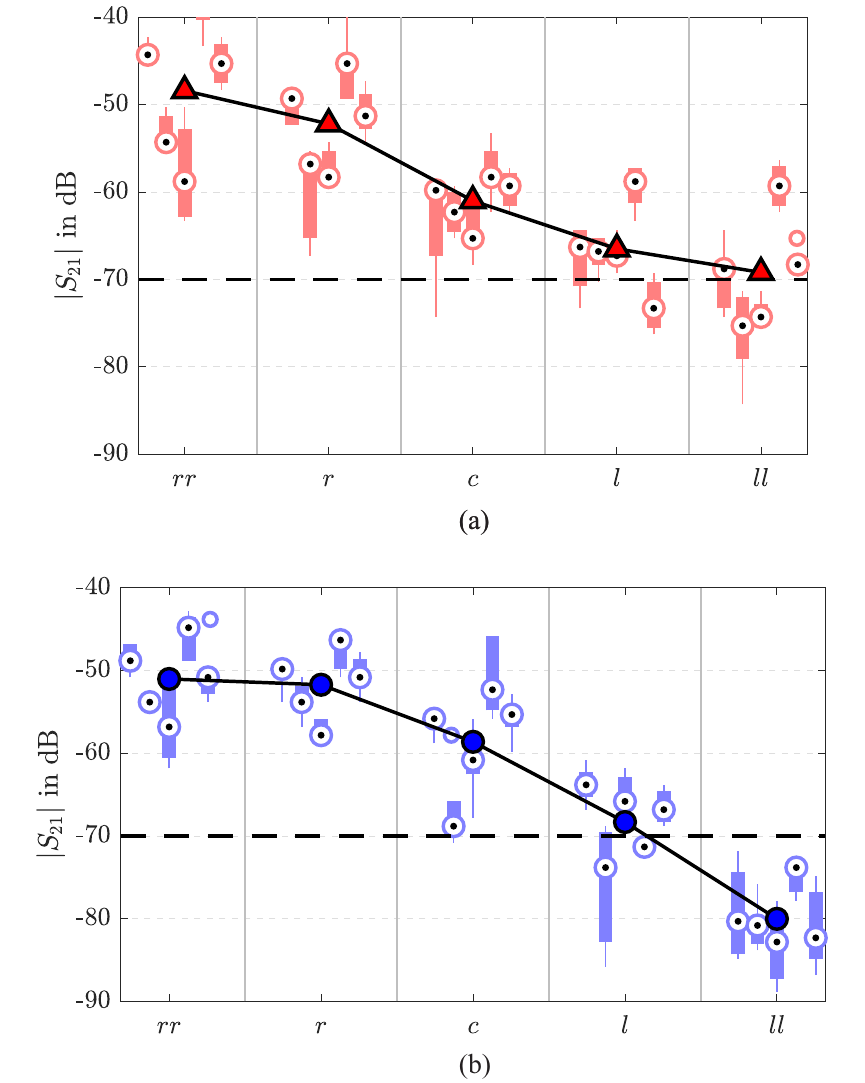}
    \caption{Distribution of the measured transmission coefficients at receiver FL: (a) Magnetic dipole; (b) Folded dipole. At each body pose ($ll$-$rr$), five boxplots depict the statistical distribution of the results of the individual measurements with the different test subjects. The calculated average transmission coefficient per pose is depicted by the bold markers. The dashed line outlines the limit for the calculation of the KPI.}
    \label{fig:example_meas_boxplot_vs_means}
\end{figure}

To further evaluate the antenna performance, the probability of connection losses is identified as a simplified key performance indicator (KPI) for the considered application. Assuming a common value for the maximum dynamic range of 90\,dB of the RF link (receiver sensitivity of -90\,dBm and transmit power of 0\,dBm), and the measured deviation of  individual values from the average of up to 20\,dB, an average transmission coefficient $|S_{21}|<-70$\,dB can be assumed to potentially cause connection losses. Accordingly, assuming that all poses are equally likely, the KPI can be calculated as the percentage of poses with $|S_{21}|<-70$. In Fig.~\ref{fig:example_meas_boxplot_vs_means}, this lower limit for the average transmission coefficient is outlined by a dashed line. If we compare the results with the two different antennas as depicted in Fig.~\ref{fig:example_meas_boxplot_vs_means} in this regard, pose $rr$ represents such a case in which with the folded dipole antenna the average transmission coefficient drops below the -70\,dB limit and also individual measured values are very close to the sensitivity limit. However, with the magnetic dipole antenna, which was designed based on the optimization, the average is increased by about 10\,dB, whereby the connection can be assumed to remain stable.

\subsection{{Evaluation}}
\begin{figure}
    \centering
    \includegraphics[width=0.95\columnwidth]{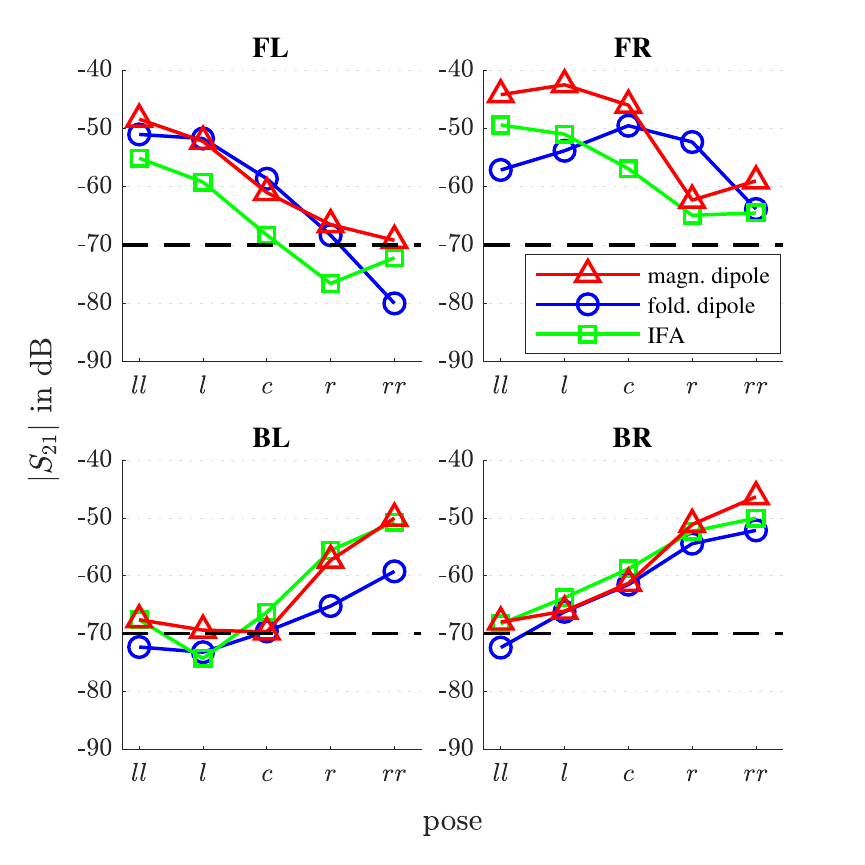}
    \caption{Measurement: average transmission coefficient in dependency of the body pose. The dashed line outlines the limit for the calculation of the KPI.}
    \label{fig:meas_all_atennas}
\end{figure}

\begin{figure}
    \centering
    \includegraphics[width=0.95\columnwidth]{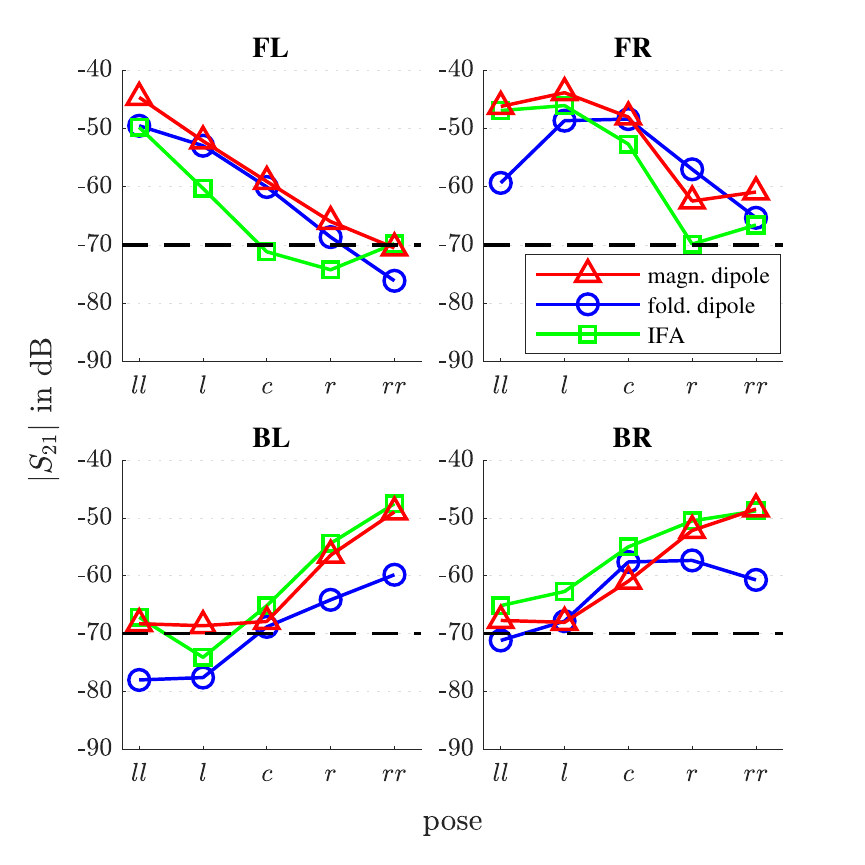}
    \caption{SWF modeling: average transmission coefficient in dependency of the body pose. The dashed line outlines the limit for the calculation of the KPI.}
    \label{fig:swf_all_atennas}
\end{figure}

Fig.~\ref{fig:meas_all_atennas} compares the measured average transmission coefficients of all three antennas considered. As can be seen, significant differences between the three antennas exist, even if the general behavior concerning the posing is similar.
The resulting average transmission coefficients derived using the SWF modeling are depicted in Fig.~\ref{fig:swf_all_atennas}. As can be seen, the main trends compared to those from the measurement in Fig.~\ref{fig:meas_all_atennas} are clearly the same. With the magnetic dipole antenna, the highest average transmission coefficients are found and thus the optimization goal is fulfilled. Depending on whether SWF modeling or measurement is considered, there is only one pose with a risk of connection losses. 

The folded dipole antenna shows the worst performance in the calculated and measured transmission coefficients. Interestingly, this antenna also shows a drop in the transmission factor to the receiver FR when the transmitter and receiver are on the same side of the body (pose $ll$). This behavior equally occurs in the measurement and the SWF modeling. In both measurement and SWF modeling, there are four poses with potential signal losses with the folded dipole. 

The transmission coefficients with the IFA are noticeably low for the receivers at the front (FL, FR), whereas at the back (BL, BR) it partly even outperforms the magnetic dipole antenna. Related to poses with potential signal losses, three cases are counted in the results of the measurement and SWF modeling with the IFA, while one case is counted at different positions in SWF modeling and measurement. In Tab.~\ref{tab:KPI}, the results of the calculated KPIs are summarized for all three antennas. For the calculation of the relative probability of connection losses, the total number of 20 equally likely poses and receiver locations is considered here.
The comparison by means of the more complex analysis of the curves in Fig.~\ref{fig:meas_all_atennas} and Fig.~\ref{fig:swf_all_atennas} can thus be translated into a quantitative factor. Its calculation with SWF modeling purely based on numerical near-field data is very fast compared to time-consuming measurements or conventional scenario-specific simulations of the entire system, with the obtained results showing very good agreement.


\begin{table}
\begin{threeparttable}[b]
\caption{Antenna Performance Evaluation}
\begin{tabular}{|p{60pt}|c|c|c|}
ine
 & magnetic dipole & folded dipole & IFA \\
 ine
KPI (SWF)\tnote{1} & 5\% (1) &  20\% (4) & 15\% (3) \\
KPI (meas.)\tnote{1} & 0\% (0) & 20\% (4) & 15\% (3) \\
 ine
\end{tabular}
\label{tab:KPI}
\begin{tablenotes}
\item [1] The KPI was calculated as the percentage of poses with potential connection losses, lower is better. The value in brackets indicates the absolute number of cases.
\end{tablenotes}
\end{threeparttable}
\end{table}

\section{Conclusion}
{
Without antenna de-embedding, WBAN systems can only be characterized and optimized as a whole including transmitter, on-body channel, and receiver. Therefore, in this contribution SWF modeling is utilized for WBAN antenna de-embedding and for deriving optimal antenna characteristics.
The issue of antenna de-embedding for WBAN, which is a considerable challenge even with SWF modeling due to the near-field coupling between the tissue and the antenna, is solved by a new straightforward approach.
Once the channel matrices have been determined, the optimal superposition of SWF modes that maximizes the incident power at the receiver can be determined analytically. This information can be used as a basis for an optimized antenna design. Also, it is possible to calculate an optimal on-body radiation pattern, as a measure for the directional radiation properties of the antenna for an on-body channel, to further support the antenna design.

Typical factors influencing WBAN channels (different body anatomies, body postures, and varying positions of the communication nodes), are taken into account in the SWF modeling and statistically evaluated in the results. Thus, overall 60 channels were modeled for different scenarios. As was shown, using the derived methods, the performance of arbitrary antennas in all these different channels can be calculated in a single step. For this purpose, only the near field of the antenna to be embedded into the channels has to be calculated in a standard EM simulation. In comparison, with standard methods simulations for each individual scenario would have to be carried out for each antenna to be evaluated. Similarly, corresponding measurements of the entire system are time-consuming. As has been shown, the SWF modeling can also be used to directly consider the performance of the antenna in different scenarios for antenna optimization.

The validation based on measurements clearly shows that the performance differences of various antennas are correctly reproduced in the derived SWF modeling. It is also shown that, as expected, the optimized antenna performs best compared to other standard antenna designs, especially in avoiding possible connection losses. By means of an exemplary derived KPI for the probability of connection losses, the great potential of the derived methods could be shown due to the very efficient computation of arbitrary channel scenarios. SWF modeling therefore provides opportunities for very detailed analysis and evaluation of antenna performance in complex environments such as body-worn applications in WBAN.
}

\ifCLASSOPTIONcaptionsoff
  \newpage
\fi


\newpage
\bibliographystyle{IEEEtran}
\bibliography{IEEEabrv,lit}

\end{document}